\documentclass[]{spie}
\usepackage{graphicx}

\begin{document}

\begin{center}
\fontsize{16}{19.2}\selectfont
\bf
Effects of differential wavefront sensor bias drifts on high contrast
imaging \\
\fontsize{12}{14.4}\selectfont
\mdseries
\bigskip

Naru Sadakuni$^{a}$, Bruce A. Macintosh$^{b}$$^{,}$$^{c}$, David W. Palmer$^{c}$,
Lisa A. Poyneer$^{c}$, Claire E. Max$^{d}$, Dmitry Savransky$^{e}$,
Sandrine J. Thomas$^{f}$, Andrew Cardwell$^{a}$, Stephen
Goodsell$^{a}$, Markus Hartung$^{a}$,
Pascale Hibon$^{a}$, Fredrik Rantakyr\"o$^{a}$, Andrew Serio$^{a}$
with the GPI team\footnote{Please send correspondence to Naru Sadakuni at nsadakuni@gemini.edu}
 \\

\bigskip
$^{a}$Gemini Observatory, La Serena, Chile \\
 $^{b}$Kavli Institute for Particle Astrophysics and Cosmology,
 Stanford University, Stanford, CA USA\\
 $^{c}$Lawrence Livermore National Laboratory, Livermore, CA USA\\
 $^{d}$University of California, Santa Cruz, Santa Cruz, CA USA \\
$^{e}$Sibley School of Mechanical and Aerospace Engineering, Cornell
University, Ithaca, NY USA\\
$^{f}$NASA Ames Research Center, Mountain View, CA USA \\

\end{center}

\section*{ABSTRACT}

\mdseries
\fontsize{10}{11.2}\selectfont

The Gemini Planet Imager (GPI) is a new facility, extreme adaptive
optics (AO), coronagraphic instrument, currently being integrated onto
the 8-meter Gemini South telescope, with the ultimate goal of directly
imaging extrasolar planets. To achieve the contrast required for the
desired science, it is necessary to quantify and mitigate wavefront
error (WFE). A large source of potential static WFE arises from the
primary AO wavefront sensor (WFS) detector's use of multiple readout
segments with independent signal chains including on-chip
preamplifiers and external amplifiers. Temperature changes within
GPI's electronics cause drifts in readout segments' bias levels, inducing an RMS WFE
of 1.1 nm and 41.9 nm over 4.44 degrees Celsius, for magnitude 4 and
11 stars, respectively. With a goal of $<$2 nm of static WFE, these
are significant enough to require remedial action. Simulations imply a
requirement to take fresh WFS darks every 2 degrees Celsius of
temperature change, for a magnitude 6 star; similarly, for a magnitude
7 star, every 1 degree Celsius of temperature change. For sufficiently
dim stars, bias drifts exceed the signal, causing a large initial WFE,
and the former periodic requirement practically becomes an
instantaneous/continuous one, making the goal of $<$2 nm of static WFE
very difficult for stars of magnitude 9 or fainter. In extreme cases,
this can cause the AO loops to destabilize due to perceived
nonphysical wavefronts, as some of the WFS's Shack-Hartmann quadcells
are split between multiple readout segments. Presented here is GPI's
AO WFS geometry, along with detailed steps in the simulation used to
quantify bias drift related WFE, followed by laboratory and on sky
results, and concluded with possible methods of remediation. \\

\bf \noindent 
Keywords: \mdseries adaptive optics, high contrast imaging, Gemini
Planet Imager, GPI

% 
%%
%%%
%%%%
%%%
%%
%

\section{Introduction}
Exoplanet detection and characterization are imperative in
understanding planet and planetary system formation and
evolution. Currently, the majority of detection methods are indirect,
limited to rough characterizations of the planet’s mass and radius. In
contrast, direct imaging actually resolves and images the assumed
planet’s light, allowing explicit determination of attributes such as
spectra and planet-star separations, in turn establishing temperature
and surface gravity, and ultimately revealing atmosphere and thermal
evolution. \\ \\
Recent advances in adaptive optics, the measurement and compensation
of wavefront distortions through high frequency CCDs, i.e. 1 kHz frame
rate, and high-order deformable mirrors, i.e. 1000 or more
actuators, have made possible the development of high contrast
astronomical imaging instruments. With an expected planet/star
contrast ratio of 10$^{-6}$ to 10$^{-7}$ from 0.2-0.8 arcseconds of planet-star
separation, and a projected sensitivity to young ($<$1GYr), Jovian-mass
planets at a distance of 5-100 AU, the Gemini Planet Imager
(GPI)\cite{Macintosh12052014} will be able to contribute greatly to the unexplored range of possible
exoplanet discovery.\\ \\
Understanding and quantifying sources of wavefront error
within GPI, discussed here, is necessary to achieving science images capable of
planet detection.

\section{AOWFS}
\subsection{CCID-66}

\mdseries
The Lincoln Labs CCID-66, used in the Shack-Hartmann wave front sensor
(SHWFS) in the adaptive optics (AO) system, consists of 160x160 active
pixels, separated into 16x64 pixel segments. The CCID-66 incorporates a planar JFET
first-stage amplifier in each segment to provide on-chip gain and
reduce readout noise, but this may contribute to increased drifts in
bias or gain levels. For GPI's purposes, only the central 128x128
pixels are used.

\subsection{Geometry}
\subsubsection{128x128}
The 128x128 pixel array used by the SHWFS is partitioned into 2x2
pixel blocks, called quadcells, by designating every third column and
row of pixels, starting from the bottom-left side of the image, as
unused bands, Figure \ref{biasx}a. These bands prevent
light from spilling into neighboring quadcells in the case of sudden,
large turbulence, hence named guard bands. The array is physically
situated such that when illuminating the lenslet array, prior in the
optical path, the resulting focal points hit said quadcells, and allow
for centroid calculations and in turn phase reconstructions. With this
configuration, the available 128x128 pixels allows a maximum 43x43
usable lenslets.\\
\noindent
Furthermore, the array is sectioned into 16x64 pixel segments, each of
which is read out by an independent signal chain including on-chip
preamplifiers and external amplifiers, Figure
\ref{biasx}a. Multiple signal chains allow for faster readout speeds
albeit, in practice, introduce discontinuities among intrinsic bias
levels of individual segments. This can lead to misleading centroid
calculations and phase reconstructions, consequently a larger
wavefront error, and ultimately issues in the science, see Section 3. 
\subsubsection{96x96}
\begin{figure} [!h]
\begin{center}
\includegraphics[width=4.5in]{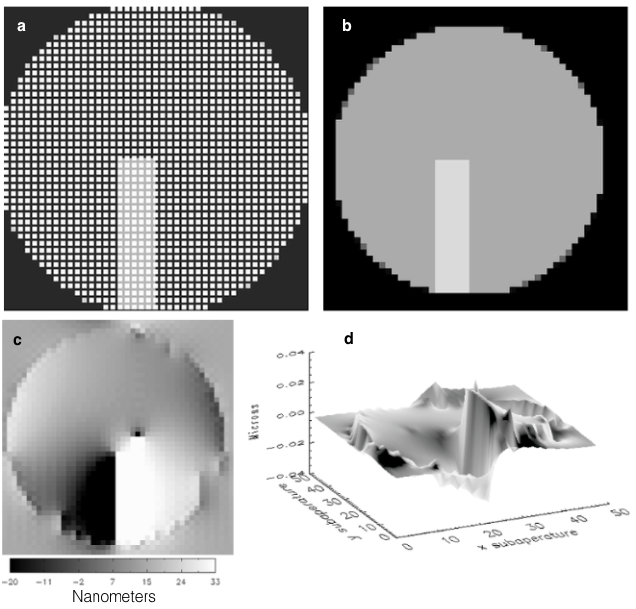}  
\caption{Major steps in the simulation are illustrated in this extreme
  case. a) The highlighted region shows a 30\% bias drift of the
  original signal in one of the sixteen 16x64 pixel
  segments with its own readout electronics. Note the unused, single
  pixel bands between quadcells, known as guard bands. Also note the
  particular geometry results in splitting of certain quadcells among
  various readout segments. b) The 96x96 array after removing guard bands. Note the 4 pixel border along the bottom
  and left as well as the 6 pixel border along the top and right,
  leaving 86 valid pixels across, or 43 quadcells across,
  corresponding to the 43 subaperatures or lenslets across. c) The phase
  reconstructed in 2-D, and d) the same phase reconstructed in 3-D.}    
\label{biasx}
\end{center}
\end{figure}

The read out 128x128 arrays have a stored bias frame subtracted, their
guard bands removed, and are manipulated such that a 96x96 array
remains consisting purely of quadcells - using the 96x96 array space,
centroid and phase reconstructions are computationally more
convenient. As previously stated, the SHWFS has 43 usable lenslets
across, leaving zero padding around the borders, Figure \ref{biasx}b.
\subsubsection{48x48}
GPI's wavefront Fourier-transform reconstructor (FTR)
algorithm\cite{poyneer} converts the 96x96 array into a 48x48 array
phase map. From a computational point of view, it is
more convenient to produce an array of these dimensions, regardless
of only using 43x43 lenslets, as it is
consistent with simulations and it leaves an array large enough to add
2 rings of MEMS slaves. In steps, the centroids are computed from the 96x96 array,
reference centroids are subtracted, and the phase is reconstructed. 
\newpage
\section{Bias drift}
\begin{figure}[!h]
\begin{center}
\includegraphics[width=5.5in]{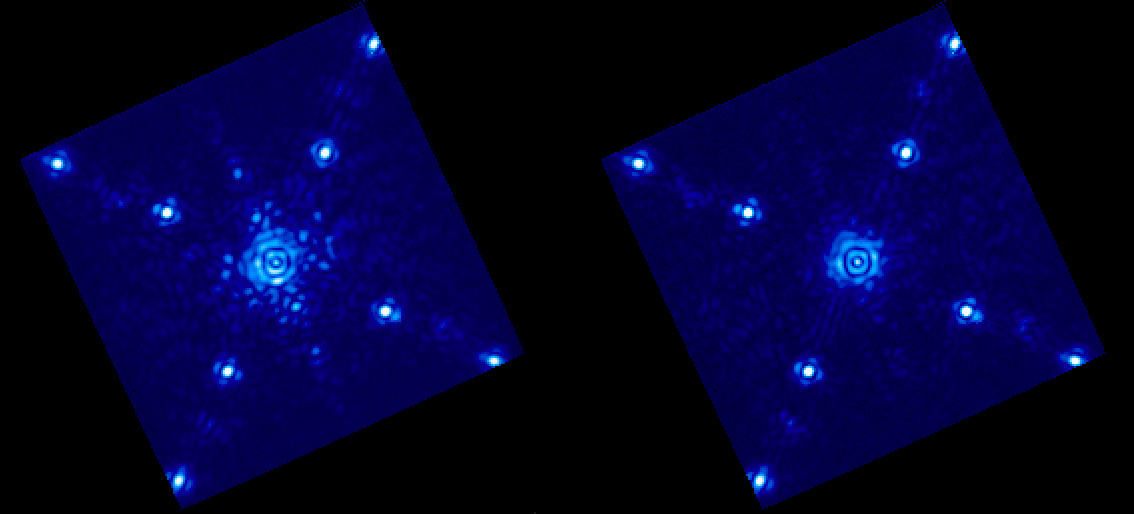}  
\caption{Showing a single wavelength channel of two different end-to-end
  images, one before (left) and one after (right) taking a new AO WFS dark, as
  well as new reference centroids. A clear improvement of the image is
seen.}    
\label{beforeandafter}
\end{center}
\end{figure}

Partitioning the 128x128 space into 16x64 pixel segments allows faster
readout speed through the use of multiple taps, one for each
segment. As a consequence, each segment has its own intrinsic bias
level, different from others'. Furthermore, each segment's bias level
fluctuates depending heavily on temperature, primarily the temperature of
GPI's electronic enclosure (EE) box containing the SHWFS electronics. This poses a problem for centroid
calculations of light focused on quadcells split between readout
segments; a difference in bias levels between two halves of a
quadcell can be misinterpreted as a slope of the wavefront, at which
point the AO will mistakenly move a DM actuator in attempt to correct
for this. With a large enough temperature change, this effect could
potentially induce significant WFE, Figure \ref{beforeandafter}.

\subsection{Simulation}
In studying the effects of bias drifts through simulation, it is
necessary to accurately replicate GPI's SHWFS geometry, described in
Section 2.2, to properly reconstruct meaningful centroids and
phases. Starting in the 128x128 space, all pixels in all valid
quadcells are assigned a specific digital number (DN). Of importance,
here, is the particular DN assigned to said pixels, as bias
fluctuations relative to the signal are ultimately of
significance. Simulating real signals, as opposed to arbitrarily
choosing DNs, is clearly more informative and thus desirable. Defined
by GPI's optics, a simple equation allows us to convert star
magnitudes to DN/ms/pixel seen by the WFS,

\begin{equation}
DN = 10^{-1[\frac{m-4}{2.5}-\log(1485)]}
\label{dn2mag}
\end{equation}
where $m$ is the user-defined star magnitude. It follows that a
dimmer star will yield less DNs and therefore a given bias change will
induce relatively more WFE. \\

\noindent
Once the star magnitude is fixed, different bias levels can be set to
individual readout segments, before removing guard bands and
converting to the 96x96 pixel space. Shown in Figure \ref{biasx}a,
some quadcells are split among various readout segments - the
boundaries of the readout segments are defined in the 128x128 pixel
space, therefore it is crucial to apply the various bias levels
here. Centroids for each quadcell in the 96x96 array are then
calculated, from which the phase is reconstructed and stored in a
48x48 array. Figure \ref{biasx} shows an extreme case where one
readout segment's bias increases by 30\% of the signal.\\ \\ 
It is more useful, once again, to apply real bias levels and drifts
seen by GPI. Over several nights bias levels across the WFS chip were
read out and recorded, in addition to numerous temperatures measured
by different sensors within GPI. These biases were then applied
accordingly, with different magnitude stars set as the signal, and
resulting phases were reconstructed. With a 4.44$^\circ$C change in
GPI's EE air inlet temperature over $\sim$12 hours, a maximum bias
drift of 95.5 DN was seen. For a magnitude 4 star, the former effect
will result in 1.074 nm RMS WFE, alternatively  0.273 nm RMS/$^\circ$C
in the linear region, justifiably adequate to leave unaccounted for.\\

\begin{figure}[!h]
\hspace{1.25in}
\includegraphics[width=5.in]{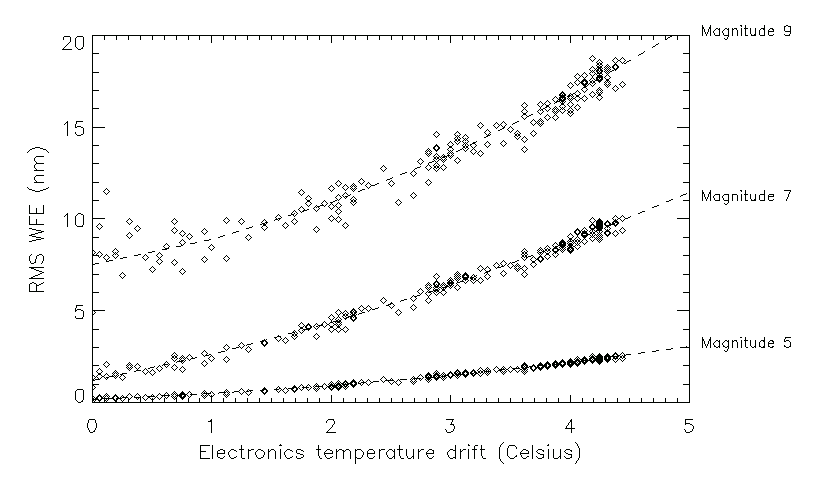}  
\caption{Showing the WFE as a function of
  relative temperature change. For magnitudes 5, 7, and 9 stars, a
  4.44$^\circ$C change causes 2.56, 10.02, and 18.73 nm RMS WFE, respectively.}    
\label{biasdriftall}
\end{figure}

\begin{figure} [h!]
\begin{center}
\includegraphics[width=5.in]{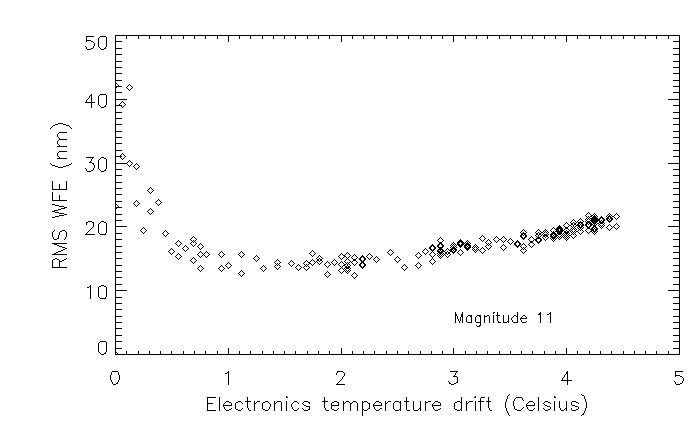}  
\caption{Showing the RMS of the WFE as a function of relative temperature change for a magnitude 11 star. Due to the weak signal, even slight variations in temperature lead to bias drifts relatively large enough to compose a significant fraction of the original signal, ultimately leading to a large, initial RMS WFE. Eventually the bias drifts sufficiently to practically overtake the signal and phase reconstructions essentially represent only bias. }    
\label{rmsmag11}
\end{center}
\end{figure}

\noindent
In comparison, when observing dimmer targets this effect becomes
significant enough to require correction. The same temperature change
ultimately results in RMS WFEs of 9.802 nm, 14.63 nm, 18.27 nm, 23.57
nm, and 41.91 nm for magnitudes 7, 8, 9, 10, and 11 stars,
respectively; correspondingly, this translates to 2.05, 2.55, 2.57,
2.40, 2.27 nm RMS/$^\circ$C in the linear region. However, for star
magnitudes $>$9 the signal seen by GPI's WFS becomes weak enough to the point where bias
drifts resulting from even fractions of a degree of temperature change
become a significant percentage of the original signal, leading to a
large, initial RMS WFE, Figure ~\ref{rmsmag11}. Eventually, the bias
drifts sufficiently as to exceed the original signal and reconstructed
phases practically represent only bias. Therefore, to mitigate WFE,
the rate at which one must update the subtracted WFS dark image
increases, not only as the rate of temperature change increases, but
also as the measured starlight decreases. 

%
%%
%%%
%%%%
%%%
%%
%
\subsection{On sky performance}
\begin{figure} [!h]
\begin{center}
\includegraphics[width=5.in,height=3.in]{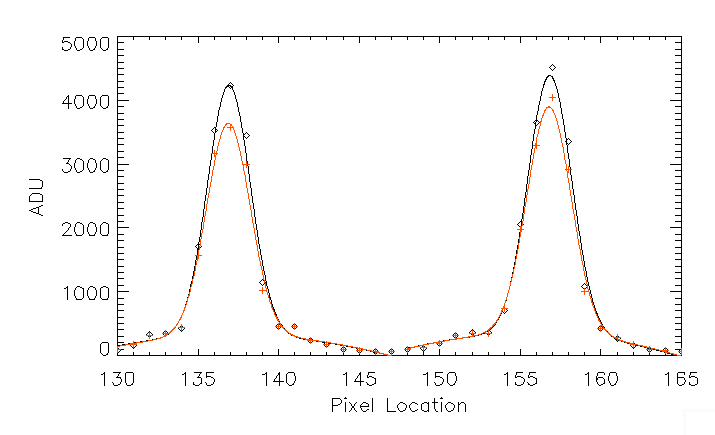}  
\caption{HD 139498 (I mag = 8.9) and its binary companion were directly imaged in H
  band and Gaussians were individually fit to a column slice through each star's PSF. With an
average bias drift of 4.59 DN, a decrease in peak counts of 18.1\% and
14.0\% are seen respectively for each star. The red plot corresponds to images
that contain static WFE from AO WFS bias drifts, while the black plot
corresponds to images that presumably contain little, if any, static
WFE from AO WFS bias drifts, or lack thereof.}    
\label{onskies}
\end{center}
\end{figure}

In order to see the effects of the bias drifts on image quality on sky, the bias was
intentionally allowed to drift while spectroscopic H band images were
taken on GPI's science detector\cite{larkin}. After a series of images, a fresh
AO WFS dark image was taken and another set of images queued. Figure~\ref{onskies} depicts column slices through the pair of exposures just before and after
the new AO WFS dark image was taken. With an average drift of 4.59 DN
on an I mag = 8.9,
the peak counts in each star of the binary system HD 139498 decreased
by 18.1\% and 14.0\% respectively; with this bias drift, simulations
imply an induced static WFE of 6.64 nm RMS. 

\section{Methods of remediation} 
\begin{figure} [!h]
\begin{center}
\includegraphics[width=5.in]{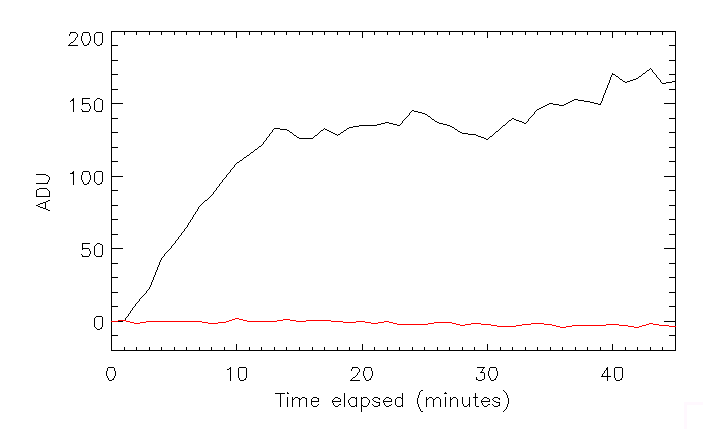}  
\caption{The unilluminated corners of the WFS are used to track the
  bias drifts and subtract them from active pixels. Here, an
  intentional temperature drift was induced, showing a clear
  effectiveness of the bias correction (red) compared to no bias
  correction (black).}
\label{autobias}
\end{center}
\end{figure}

\begin{figure} [!h]
\begin{center}
\includegraphics[width=5.in]{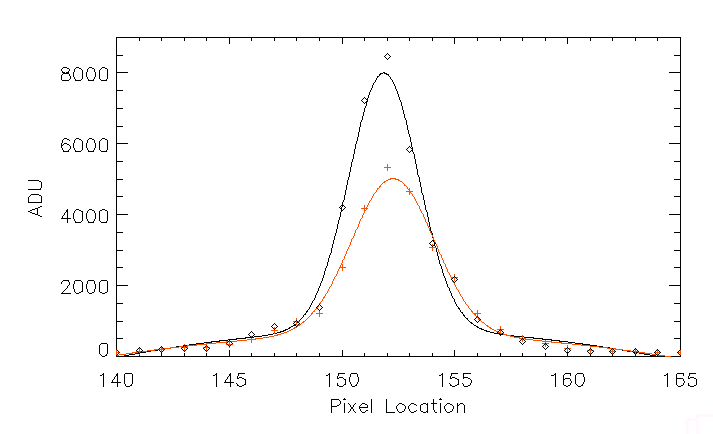}  
\caption{HD 76608 (I mag = 9.9) was directly imaged in H band and a Gaussian fit to
  a column slice through the PSF. With bias drift compensation turned on (black), the FWHM decreases
  by 17\% and the peak pixel count increases by 33\% compared to
  without bias drift compensation (red).}
\label{darkcorner}
\end{center}
\end{figure}

A method of remediation that has been implemented and tested, is the use of the
unilluminated corner pixels of the WFS to track the bias drifts. The
DN of each corner pixel is boxcar averaged, then averaged with each other, determining a single
correction value. Said correction value is differenced from the dark
value and applied to all active pixels in every frame. Figure
~\ref{autobias} shows the effectiveness of the correction over a
period of 45 minutes. However, this correction does not mitigate frame to frame
fluctuations, nor does it correct for readout segment differential
bias drifts. This method compensates for nondifferential bias
drifts across the WFS, which also have significant effects if left
uncorrected. Take, for example, centroid calculations for the simplified case of a 1-D
''quadcell'' with just 2 pixels of intensities A and B: 
\begin{equation}
Center of Gravity = \frac{(A-B)}{(A+B)}
\label{centroid}
\end{equation}
If a uniform bias drift, N, is applied to the entire ''quadcell'', the
measured centroid becomes:
\begin{equation}
Center of Gravity = \frac{(A+N)-(B+N)}{(A+N)+(B+N)} = \frac{(A-B)}{(A+B+2N)}
\label{centroid1}
\end{equation}
For N$>$0, the centroider will report a lower centroid, leading to
underestimation of the phase and hence undercorrection. Ultimately,
the AO will converge to the right solution, as it is running in closed
loop, but effectively the centroid and control loop have a lower gain,
therefore reducing the temporal bandwidth for correcting atmospheric
turbulence. Furthermore, if a nonzero reference centroid exists that
corresponds to a particular, correct actuator position/local slope,
the centroider will not measure the same value at times when it
should, and will have to overcorrect. Thus, this method of correction
remains useful and has proven effective on sky
by comparing end to end science images, Figure ~\ref{darkcorner}.

\newpage
\section{Future work}
A method of remediation in consideration for the future is the use of the
unilluminated top and bottom pixel rows of the CCID-66 to track the
differential bias drifts. With this method it would be possible to
correct the bias drifts individually for each segment. The bias drifts
explained in this paper were unforeseen during design stages, thus, for
increased speed, the
top and bottom rows are currently not read out, and therefore this
method could not yet be implemented nor tested.

\section{Acknowledgements}
The Gemini Observatory is operated by the Association of Universities
for Research in Astronomy, Inc., under a cooperative agreement with
the NSF on behalf of the Gemini partnership: the National Science
Foundation (United States), the National Research Council (Canada),
CONICYT (Chile), the Australian Research Council (Australia),
Ministerio da Ciencia, Tecnologia e Inovacao (Brazil), and Ministerio
Innovacion Productiva (Argentina). We acknowledge financial support of
the Gemini Observatory, the Center for Adaptive Optics at UC Santa
Cruz (NSF AST-9876783), the NSF (AST-0909188; AST-1211562), NASA
Origins (NNX11AD21G; NNX10AH31G), the University of California Office
of the President (LFRP-118057), and the Dunlap Institute, University
of Toronto.

\bibliography{refs}
\bibliographystyle{spiebib}

\end{document}